\documentclass[preprint,showpacs,preprintnumbers,amsmath,amssymb]{revtex4} 
 
\usepackage{graphicx}
\usepackage{dcolumn}
\usepackage{bm}
 
\begin{document} 
 
\title{Effective photon mass in nuclear matter and finite nuclei} 
 
\author{Bao-Xi Sun$^{1,2}$, Xiao-Fu Lu$^{2,3,6}$, Peng-Nian Shen$^{6,1,2}$, 
En-Guang Zhao$^{2,4,5,6}$} 
 
\affiliation{${}^{1}$Institute of High Energy Physics, The Chinese Academy of 
Sciences, P.O.Box 918(4), Beijing  100039, China} 
 
\affiliation{${}^{2}$Institute of Theoretical Physics, The Chinese Academy of 
Sciences, Beijing  100080, China} 
 
\affiliation{${}^{3}$Department of Physics, Sichuan University, 
Chengdu  610064, China} 
 
\affiliation{${}^{4}$Center of Theoretical Nuclear Physics, 
National Laboratory of Heavy ion Accelerator,Lanzhou 730000,  China} 
 
\affiliation{${}^{5}$Department of Physics, Tsinghua University, 
Beijing 100084, China} 
 
\affiliation{${}^{6}$China Center of Advanced Science and Technology(World 
Laboratory), Beijing 100080, China} 
 
\begin{abstract} 
Electromagnetic field in nuclear matter and nuclei are studied. 
In the nuclear matter, because the expectation value of the electric charge 
density operator  is not zero, different in vacuum, the 
$U(1)$ local gauge symmetry of electric charge is spontaneously 
broken, and consequently, the photon gains an effective mass through 
the Higgs mechanism. An alternative way to study the effective mass 
of photon is to calculate the self-energy of photon perturbatively. 
It shows that the effective mass of photon is about $5.42MeV$ 
in the symmetric nuclear matter 
at the saturation density $\rho_0~=~0.16fm^{-3}$ 
and about $2.0MeV$ at the surface of ${}^{238}U$. It seems that 
the two-body decay of a massive photon causes the sharp lines 
of electron-positron pairs in the low energy heavy ion 
collision experiments of ${}^{238}U+{}^{232}Th$ . 
\end{abstract} 
 
\maketitle 

\newpage       

\section{Introduction}

The observation of electron-positron pairs in high-Z heavy ion
collision experiments at lower energies has been paid close
attention \cite{cowa/86,cowa/87,sala/90}. From the Monte Carlo
simulations of the lepton kinetic-energy and time-of-flight
distributions,  it seems that there exists  a prompt two-body
decay of a light neutral particle of mass $1.8MeV$ at rest in the
center-of-mass frame. Although it was claimed that the sharp
sum-energy lines in electron-positron pairs emission from
heavy-ion collisions was not repeated  by the ATLAS Positron
Experiment$(APEX)$ in 1995 \cite{APEX95}, the APEX results have
been suspected \cite{Gri96,Gri9706,Gri9741,Gri98}. In this paper,
by studying the properties of the electromagnetic field in the
nuclear matter and finite nuclei, we would show that the predicted
light neutral particle of mass $1.8MeV$ is just a massive photon.
In nuclear matter, we suggested that the photon gains mass through
the interaction with nucleons, and the same mechanism may also be
applied to the electromagnetic field in finite nuclei.

\section{Perturbation calculation of effective photon mass in nuclear matter}

In nuclear physics, we always define the ground state of nuclear
matter as "vacuum", where the Fermi sea is filled by nucleons, and
no anti-nucleons and holes exist.
Because the proton has one unit of positive charge,
and the expectation value of electric density operator
in the nuclear matter is not zero, namely
$\langle~\hat{\rho}_Q~\rangle~\not=~0$.
This "vacuum" is different from the real vacuum
where nothing exists.

If the electromagnetic field and proton field interaction is considered as a
perturbation,  the perturbative hamiltonian in the interactive representation
can be expressed as
\begin{equation}
{\cal H}_I~=~e\bar\psi_{p}(x)\gamma^{\mu}\psi_{p}(x)A_{\mu}(x),
\end{equation}
with
$\psi_p$ being the fields of the proton,
and $A_\mu$ being the electromagnetic field.

The S-matrix can be written as
\begin{equation}
\hat{S} ~=~\hat{S}_0~+~\hat{S}_1~+~\hat{S}_2~+~\ldots,
\end{equation}
where
\begin{equation}
\hat{S}_n ~=~\frac{(-i)^n}{n !}\int d^{4}x_1 \int d^{4}x_2 \ldots \int
d^{4}x_n   T \left[{\cal H}_I(x_1){\cal H}_I(x_2) \ldots {\cal H}_I(x_n)
\right].
\end{equation}
To study the proton-proton interaction via the photon exchange,
\begin{equation}
\label{eq:S2}
\hat{S}_2 ~=~\frac{(-i)^2}{2 !}\int d^{4}x_1 \int d^{4}x_2
 T\left[{\cal H}_I(x_1){\cal H}_I(x_2) \right].
\end{equation}
should be calculated.

The proton field operator $\psi_p(x)$ and its conjugate operator
$\bar\psi_p(x)$ can be expanded in terms of a complete set of solutions to
the Dirac equation:
\begin{equation}
\label{eq:pfield}
\psi_p(x)~=~\sum_{\lambda=1,2}\int\frac{d^3p}{(2\pi)^\frac{3}{2}}
\sqrt{\frac{m}{E(p)}}A_{p,\lambda}U(p,\lambda)\exp(-ip_{\mu}x^{\mu}),
\end{equation}
\begin{equation}
\label{eq:bpfield}
\bar\psi_p(x)~=~\sum_{\lambda=1,2}\int\frac{d^3p}{(2\pi)^\frac{3}{2}}
\sqrt{\frac{m}{E(p)}}A^{\dagger}_{p,\lambda}\bar{U}(p,\lambda)\exp(ip_{\mu}x^{\mu}),
\end{equation}
where $E(p)= \sqrt{\vec{p}^2+m^2}$, and $\lambda$ denotes the spin
of the proton, $A_{p,\lambda}$ and $A^{\dagger}_{p,\lambda}$ are
the annihilation and creation operators of the nucleon
respectively. In these two expressions, we have assumed that there
are no antinucleons in the nuclear matter or finite nuclei, thus
only positive-energy components exist.
The photon field operator $A_\mu(k,x)$ can be expressed as
\begin{equation}
A_\mu(k,x)~=~
a(k,\delta)\varepsilon_\mu(k,\delta)\exp(-ik \cdot x)
+a^\dagger(k,\delta)\varepsilon_\mu(k,\delta)\exp(ik \cdot x),
\end{equation}
with $k$ being the momentum of the photon.

Since a perturbative single nucleon loop would contribute to the
self energy of the photon a term of $-e_{0}^{2}k^{2}C(k^2)$, which
can be eliminated by the renormalization procedure\cite{Ru.68},
the diagrams in Fig.1 should be calculated.

The expectation value of $\hat{S}_2$ can be
written as
\begin{eqnarray}
\label{eq:S2-1}
\langle~k_2,\varepsilon_\mu(k_2,\delta_2)~|~\hat{S}_2~|~
k_1,\varepsilon_\nu(k_1,\delta_1)~\rangle
~&=&~-ie^2(2\pi)^4 \delta^4(p_1+k_1-p_2-k_2)
\varepsilon_\mu(k,\delta)~\varepsilon_{\nu}(k,\delta) \nonumber\\
&&\sum_{\lambda=1,2}\int\frac{d^3p}{(2\pi)^3}
\frac{m}{E(p)}\theta(p_F-|\vec{p}|) \\
&&\bar{U}(p,\lambda)
\left(\gamma^{\nu}\frac{1}{\rlap{/} p -\rlap{/} k - m}\gamma^{\mu}~+~
\gamma^{\mu}\frac{1}{\rlap{/} p +\rlap{/} k -
m}\gamma^{\nu}\right)
 U(p,\lambda), \nonumber
\end{eqnarray}
where $k_1=k_2=k, $ and $p_1=p_2=p, $ and $\theta(x)$ is the step function.

As the situation in Fig.1 is considered, we obtains the photon
propagator $G(k)$ in
nuclear matter as
\begin{eqnarray}
G(k)~&=&~\frac{1}{(2\pi)^4}\frac{-ig_{\mu\nu}}{k^2+i\varepsilon}+
\frac{1}{(2\pi)^4}\frac{-ig_{\mu\alpha}}{k^2+i\varepsilon}
\sum_{\lambda=1,2} (-ie^2)(2\pi)^4 \int\frac{d^3p}{(2\pi)^3}
\frac{m}{E(p)}\theta(p_F-|\vec{p}|) \nonumber \\
&&\bar{U}(p,\lambda)
\left(\gamma^{\beta}\frac{1}{\rlap{/} p -\rlap{/} k - m}\gamma^{\alpha}~+~
\gamma^{\alpha}\frac{1}{\rlap{/} p +\rlap{/} k -
m}\gamma^{\beta}\right)
 U(p,\lambda)
\frac{1}{(2\pi)^4}\frac{-ig_{\beta\nu}}{k^2+i\varepsilon} \nonumber \\
~&=&~\frac{-i}{(2\pi)^4}
\frac{g_{\mu\nu}}{k^2+i\varepsilon}+\frac{-i}{(2\pi)^4}
\frac{-e^2}{k^2+i\varepsilon}
\sum_{\lambda=1,2} \int\frac{d^3p}{(2\pi)^3}
\frac{m}{E(p)}\theta(p_F-|\vec{p}|) \nonumber \\
&&\bar{U}(p,\lambda)
\left(\gamma_{\nu}\frac{1}{\rlap{/} p -\rlap{/} k - m}\gamma_{\mu}~+~
\gamma_{\mu}\frac{1}{\rlap{/} p +\rlap{/} k -
m}\gamma_{\nu}\right)
 U(p,\lambda)
\frac{1}{k^2+i\varepsilon}.
\end{eqnarray}
According to the Dyson equation
\begin{equation}
\frac{-ig_{\mu\nu}}{k^2-\mu^2+i\varepsilon}
~=~\frac{-ig_{\mu\nu}}{k^2+i\varepsilon}
+\frac{-i}{k^2+i\varepsilon} ~ g_{\mu\nu}\mu^2 ~
\frac{1}{k^2+i\varepsilon},
\end{equation}
we obtain
\begin{eqnarray}
g_{\mu\nu}~\mu^2~&=&~
-e^2~\sum_{\lambda=1,2} \int\frac{d^3p}{(2\pi)^3}
\frac{m}{E(p)}\theta(p_F-|\vec{p}|) \nonumber \\
&&\bar{U}(p,\lambda)
\left(\gamma_{\nu}\frac{1}{\rlap{/} p -\rlap{/} k - m}\gamma_{\mu}~+~
\gamma_{\mu}\frac{1}{\rlap{/} p +\rlap{/} k -
m}\gamma_{\nu}\right)
 U(p,\lambda).
\end{eqnarray}

Under the on shell condition of protons
\begin{equation}
p^2-m^2~\approx~0,
\end{equation}
the self-energy of the real, on-shell photon
can be easily derived as
\begin{eqnarray}
\mu^2~&=&~\frac{e^2}{m}\int\frac{d^3p}{(2\pi)^3}
\frac{m}{E(p)}\theta(p_F-|\vec{p}|) \nonumber\\
~&=&~\frac{e^2 \rho_S^p}{2m}.
\end{eqnarray}
In this expression, $e^2~=~4\pi \alpha$,
with $\alpha=\frac{1}{137}$ being the fine structure constant,
$m$ is the mass of proton, and $\rho_S^p$ denotes
the scalar density of protons
\begin{equation}
\rho_S^p=
 2 \int_{p}\frac{d^3p}{(2\pi)^3}\frac{m}{(\vec{p}^2+m^2)^\frac{1}{2}}.
\end{equation}
Thus, the photon gains the effective mass
\begin{equation}
\mu~=~\sqrt{\frac{e^2 \rho_S^p}{2m}}
\end{equation}
in nuclear matter.

It should be mentioned that
the effective mass of photon is only related to the
scalar density of protons in the
nuclear matter, but not the momentum of the photon.
In a symmetric nuclear matter, where the densities of proton and
neutron are the same,
If the
nucleon density is $0.16fm^{-3}$, the effective mass of photon
is about $5.42MeV$.

In a finite nucleus, a two-parameter Fermi model of nuclear
charge-density-distribution can be written as\cite{VR.87}:
\begin{equation}
\rho_p(r)~=~\frac{\rho_c}{1+\exp(\frac{r-c}{z})},
\end{equation}
where $c=6.874fm$, $z=0.556fm$ for ${}^{238}U$, and
$\rho_c=0.0635fm^{-3}$ is determined by the equation
\begin{equation}
4\pi\int\rho_p(r) r^2 dr~=~Z.
\end{equation}
By assuming that in a finite nucleus, the
proton number density is equal to the nuclear charge-density,
the proton densities and corresponding effective photon masses at
different radii around the surface region of ${}^{238}U$
can be calculated and the corresponding results
 are listed in Table
1. It seems that at the surface of ${}^{238}U$, the photon has
effective mass of about $2.0MeV$, which is consistent with the
predicted neutral particle mass in the low energy
${}^{238}U+{}^{232}Th$ heavy-ion collisions near Coulomb barrier
\cite{cowa/86,cowa/87,sala/90}, so the massive photon at the
surface of ${}^{238}U$ or ${}^{232}Th$ might decay into an
electron-positron pair in the disturbances of the nuclear Coulomb
field. It might be the reason of the discovery of sharp line
800keV $e^{+}e^{-}$ pairs in the ${}^{238}U+{}^{232}Th$ heavy-ion
collisions.

The electromagnetic interaction between protons is realized by
exchanging virtual photons in nuclear matter. In the relativistic
Hartree approximation shown in Fig. 2, the momentum of
virtual proton $k=0,$ consequently $k^2=0$. Then the argument for the real
photon mentioned above is suitable for the virtual photon exchanged
between protons
in nuclear matter. The contribution of the photon mass term should
be included in the calculation of relativistic Hartree
approximation or relativistic mean-field approximation in the
finite nuclei.

\section{Higgs mechanism}

Again in the nuclear matter, because
$\langle~\rho_{p}~\rangle~\not=~0$, the $U(1)$ local gauge
symmetry of electric charge is spontaneously broken, and the Higgs
field should be included.

The Lagrangian density of Higgs field $\phi$ is\cite{Huang}
\begin{equation}
{\cal L}=
(D^\mu\phi)^{\ast} (D_\mu\phi)~-~V(\phi^{\ast}~\phi),
\end{equation}
with
\begin{equation}
D^\mu~=~\partial^{\mu}~+~ieA^\mu,
\end{equation}
\begin{equation}
V(\phi^{\ast}~\phi)~=~\lambda ~\left(\phi^{\ast}~\phi~-~\phi_{0}^2\right)^2
        (\phi_0\not=0).
\end{equation}

The Lagrangian is invariant under the local gauge transformation
\begin{equation}
A^\mu(x)~\longrightarrow~A^\mu(x)~+~\partial^{\mu}\omega(x),  \nonumber
\end{equation}
\begin{equation}
\phi(x)~\longrightarrow~e^{-ie\omega(x)}\phi(x), \nonumber
\end{equation}
\begin{equation}
\phi^{\ast}(x)~\longrightarrow~e^{ie\omega(x)}\phi^{\ast}(x),
\end{equation}
where $\omega(x)$ is an arbitrary real function. The local gauge
symmetry is spontaneously broken when $\phi_0~\not=~0$.

Obviously, the lowest-energy solution is
\begin{equation}
A^\mu(x)~=~0,
\end{equation}
\begin{equation}
\phi(x)~=~\phi_0~e^{i\alpha_0}.
\end{equation}
To study the classical modes near such a solution, the most
convenient way is staying in the
"unitary gauge", in which $\phi(x)$ is real
under a continuous local gauge transformation.
Thus we can always write
\begin{equation}
\phi(x)~=~\rho(x)~~~~~(real).
\end{equation}
Then the equations of motion become
\begin{equation}
\label{eq:em}
\frac{\partial F_{\mu\nu}}{\partial x_\nu}
~=~2e^2\rho^2 A_\mu
-e\bar\psi_{p}\gamma_\mu\psi_p,
\end{equation}
\begin{equation}
\label{eq:Hig}
(\partial^{\mu}~+~ieA^\mu)(\partial_{\mu}~+~ieA_\mu)\rho~=~
2\lambda\rho(\phi_{0}^2~-~\rho^2),
\end{equation}
\begin{equation}
\left(i\gamma_{\mu}D^{\mu}  - m\right)\psi_p~=~0,
\end{equation}
\begin{equation}
\left(i\gamma_{\mu}\partial^{\mu}  - m\right)\psi_n~=~0.
\end{equation}
Since $\partial_{\nu}\partial_{\mu}F^{\mu\nu}~=~0$, we must have
\begin{equation}
\partial_{\mu}A^{\mu}(x)~=~0,
\end{equation}
wherever $\rho(x)~\not=~0$. Taking
\begin{equation}
\rho(x)~=~\phi_0~+~\eta(x)
\end{equation}
and treating $\eta(x)$ and $A^{\mu}(x)$ as small quantities, one
can deduce Eqs.~(\ref{eq:em}) and ~(\ref{eq:Hig}) into linearized
forms
\begin{equation}
\label{eq:em2}
\frac{\partial F_{\mu\nu}}{\partial x_\nu}
~=~2e^2\phi_{0}^2 A_\mu
-e\bar\psi_{p}\gamma_\mu\psi_p,
\end{equation}
and
\begin{equation}
\label{eq:Hig2}
\left(\partial^{\mu}\partial_{\mu}~-~4\lambda\phi_{0}^2\right)\eta~=~0.
\end{equation}

These equations show that the fields  $A^1$, $A^2$,
$\phi$ and $\phi^\ast$ in a fixed gauge are now replaced by
$A^1$, $A^2$, $A^3$ and $\eta$. The spin-1 particle has a
mass of $\sqrt{2}e\phi_0$, and the spin-0 particle
has a mass of $2\sqrt{\lambda}\phi_0$,  In other word,
the  photon obtains mass in the nuclear matter.

From above discussion, we find that in the nuclear matter, the effective mass
of photon can either be obtained by the Higgs mechanism or the
perturbutive calculation. This result also implies that the Higgs mechanism is not a unique method to
solve problems when the local gauge symmetry is spontaneously broken.

\section{Summary}

In summary, the electromagnetic field in nuclear matter is studied.
Because in the nuclear matter the expectation value of the electric charge density operator
is not zero, the
$U(1)$ local gauge symmetry of electric charge is spontaneously broken, consequently
the photon gains a effective mass.
Alternatively, this mass can also be
obtained  from the perturbative calculation of the photon self-energy in the nuclear
matter. It is shown that the effective mass is related to the scalar
density of protons. The effective mass of photon is about $5.42MeV$
in the symmetric nuclear matter
at the saturation density of $\rho_0~=~0.16fm^{-3}$
, and  about $2.0MeV$ in the surface of ${}^{238}U$.
It just fits the mass of
the neutral particle suggested in the high-Z heavy ion collision experiments at low energies near Comloub barrier, in which
the sharp lines of electron-positron pair was observed. Obviously, the two-body decay
of such a massive photon would produce a pair of electron and positron.
The property of electromagnetic field in the nuclear matter
or finite nuclei is different from that in the
vacuum.
 Because the
photon in the nuclear matter has an effective mass , this fact should also carefully be considered
in the investigation of the nuclear many-body problems.
 In the future work, we will continue
to  study the relevant effects on this aspect.

The authors are grateful for the constructive discussions with Dr. Terry
 Goldman. This work was supported in part by the Major State Basic Research
Development Programme under Contract No. G2000-0774, the CAS
Knowledge Innovation Project No. KJCX2-N11 and the National
Natural Science Foundation of China under grant numbers 10075057
and 90103020.

\newpage

\begin{table}
\caption{\label{tab:table1} Effective photon mass $\mu$ and
corresponding radii $r$ in the nucleus of ${}^{238}U$ for
different proton number density $\rho_p(r)$} \vspace*{0.5cm}
\begin{center}
\begin{tabular}{lcr}\hline
$\rho_p(r)~(fm^{-3})$      &$r~(fm)$      &        $\mu~(MeV)$
\\ \hline
0.005         & 8.242             & 1.37     \\
0.006         & 8.131             & 1.50     \\
0.007         & 8.035             & 1.62     \\
0.008         & 7.951             & 1.73     \\
0.009         & 7.876             & 1.83     \\
0.010         & 7.807             & 1.93     \\
0.011         & 7.743             & 2.03     \\
0.012         & 7.684             & 2.12     \\ \hline
\end{tabular}
\end{center}
\end{table}

\newpage

\begin{figure*}
\includegraphics{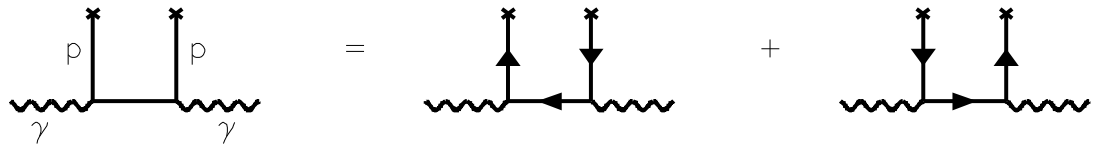}
\caption{\label{fig1}Feynman diagrams for the photon propagator in 
nuclear matter.}
\end{figure*}

\begin{figure*}
\includegraphics{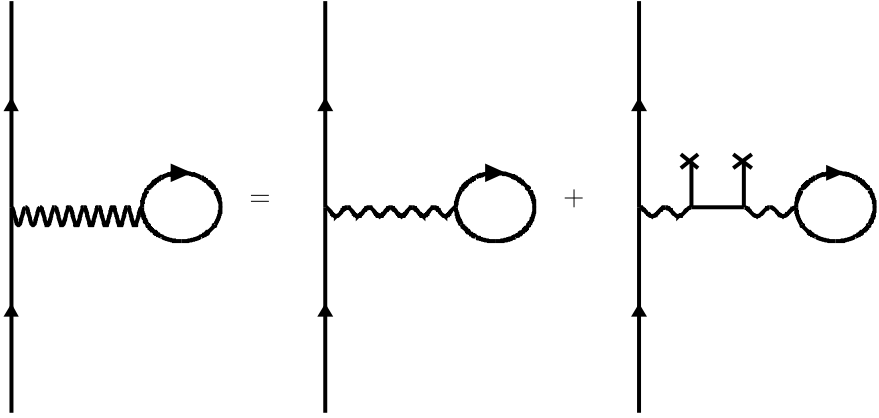}
\caption{\label{fig2}Feynman diagrams for the electromagnetic interaction 
in the relativistic Hartree approximation in finite nuclei.}
\end{figure*}

\end{document}